\documentclass[twocolumn,preprintnumbers,amsmath,amssymb,nofootinbib,epsfig,bmamsfonts,yfonts,superscriptaddress]{revtex4}
\usepackage{graphicx}
\usepackage{comment}
\usepackage[colorlinks,linktocpage,bookmarks=false,citecolor=blue,linkcolor=blue,urlcolor=blue]{hyperref}

 \def\qb{{\bar{q}}}
  \def\Hc{{\cal H}}
 \newcommand{\bk}{\mathbf{k}}
      
     \def\phic{{\phi_3}}
    \def\phid{{\phi_4}}
         \def\phib{{\phi_2}}
    \def\phia{{\phi_1}}
        \def\nus{{\nu_s}}
            \def\nuv{{\nu_\eta}}

    \def\exb{\bar{\varepsilon}}

 \def\kt{{\tilde{k}}}

  \def\phit{{\tilde{\phi}}}
   
  \def\kx{\kappa}  
 
\relax

\begin{document}

\title{Backreaction from inhomogeneous matter fields during large-scale structure formation}


\author{Stefan Floerchinger}
\email{stefan.floerchinger@thphys.uni-heidelberg.de}
\affiliation{Institut f\"{u}r Theoretische Physik, Universit\"{a}t Heidelberg, Philosophenweg 16, 69120 Heidelberg, Germany}
\author{Nikolaos Tetradis}
\email{ntetrad@phys.uoa.gr}
\affiliation{Department of Physics, University of Athens, Zographou 157 84, Greece}
\author{Urs Achim Wiedemann}
\email{urs.wiedemann@cern.ch}
\affiliation{Theoretical Physics Department, CERN, CH-1211 Gen\`eve 23, Switzerland}



\begin{abstract} 
We study how inhomogeneities of the cosmological fluid fields backreact on the homogeneous part of energy density and how they
modify the Friedmann equations. 
In general, backreaction requires to go beyond the pressureless ideal fluid approximation, and this can lead to a reduced growth of cosmological large scale structure. Since observational evidence favours evolution close to the standard growing mode in the linear regime, we focus on two-component 
fluids in which the non-ideal fluid is gravitationally coupled to cold dark matter and in which a standard growing mode persists. 
This is realized, e.g. for a baryonic fluid coupled to cold dark matter. We calculate the backreaction for this case and for a wide range of other two-fluid models. Here the effect is either suppressed because the non-ideal matter properties are numerically
too small, or because they lead to a too stringent UV cut-off of the integral over the power spectrum  that determines backreaction. We discuss then matter field backreaction from a broader perspective and generalize the formalism such that also far-from-equilibrium scenarios relevant to late cosmological times and non-linear scales can be addressed in the future.
\end{abstract}

\maketitle

\section{Introduction}
\label{sec1}

The evolution equations used in cosmology are usually obtained under the assumption of a spatially homogeneous and isotropic distribution of matter. 
Although these symmetries remain preserved at all times in a statistical sense, the deviations from homogeneity and isotropy 
become locally sizeable during cosmological large-scale structure formation at late times. Here we ask whether this can have any influence on the cosmological evolution equations.

Because Einsteins gravitational field equations are non-linear, this question amounts to asking whether
the Einstein equations for spatially averaged fields are affected by so-called 
\emph{backreaction} effects, i.e., averages of second and higher orders in spatial inhomogeneities. Arguments that such backreaction effects are always irrelevant for cosmology have been made~\cite{Green:2016cwo} and 
have been contested~\cite{Buchert:2015iva,Buchert:2011sx}, the discussion focussing mainly on inhomogeneities of the 
metric. The smallness of the gravitational coupling constant $G_\text{N}$ makes it plausible that terms non-linear in inhomogeneities of the metric
remain negligible for cosmological evolution \cite{Wetterich:2001kr, Green:2010qy}.

However, the Einstein equations contain also the energy-momentum tensor of the 
matter distribution which develops large inhomogeneities. Starting from an approximation of matter fields as a non-ideal fluid, it was demonstrated in Ref.~\cite{Floerchinger:2014jsa} how such inhomogeneities give rise to backreaction effects in late-time cosmology where non-linear terms in metric inhomogeneities and their temporal variations are negligible compared to those of inhomogeneities in the matter fields. 

The purpose of the present manuscript is 
two-fold. First, we provide within the formalism of Ref.~\cite{Floerchinger:2014jsa} the first explicit calculations  
of backreaction effects in dynamical model scenarios. Second, we generalize the formalism of  Ref.~\cite{Floerchinger:2014jsa}  to 
systems in which a Navier-Stokes fluid description cannot be taken for granted. 

\section{Backreaction formalism}
\label{sec2}
As derived in Ref.~\cite{Floerchinger:2014jsa},  backreaction effects in a Friedmann-Robertson-Walker (FRW) universe 
can be encoded in a time or scale-factor dependent source term $D(a)$ for the evolution of the spatially averaged energy density $\bar\varepsilon$,
 \begin{equation}
\tfrac{1}{a}\dot{\bar \varepsilon} + 3 H \, (\bar \varepsilon + \bar p_\text{eff}) = D(a) \label{eq1} \, .
\end{equation}
Here, the Hubble constant $H = \dot a / a^2$ is expressed in terms of the scale factor $a(\tau)$ of the 
FRW metric $ds^2 = a^2 \left[ - d\tau^2 + \delta_{ij} dx^i\, dx^j \right]$ and the dot denotes a derivative
with respect to conformal time $\tau$. The effective pressure $\bar p_\text{eff}= \bar p+\bar \pi_\text{bulk}$ contains thermal pressure and a possible bulk viscous modification. Eq.~\eqref{eq1} holds irrespective of whether a fluid description of the matter distribution with 
inhomogeneities is applicable (see section~\ref{sec4} for details), but the explicit form of $D$ depends on it. 

The spatial average of the trace of the Einstein field equations can be
shown to be free of backreaction effects, 
\begin{equation}
\frac{\ddot a}{a^3} = \frac{1}{a} \dot H + 2 H^2 = \frac{4\pi G_\text{N}}{3} \left(\bar \varepsilon - 3 \bar p_\text{eff} \right)\, .
\label{eq2}
\end{equation}
Combined with the energy conservation \eqref{eq1}, this closes the evolution equations for background fields. The two equations
combine to
\begin{equation}
	aD 
	= \frac{1}{a^4} \frac{d}{d\tau} \left[a^4 \bar\varepsilon \right] -  \frac{3}{8\pi G_N} \frac{1}{a^4}   \frac{d}{d\tau} \left[{\dot a}^2  \right]\, .
	\label{eq3}
\end{equation}
Integrating over $\tau$, one finds the modified Friedmann equation
\begin{eqnarray}
	H^2(\tau) &=& \left(\frac{{\dot a}(\tau)}{a^2(\tau)} \right)^2\label{eqn4} \\
	&=& \frac{8\pi G_N}{3} \left( \bar\varepsilon(\tau) - \frac{1}{a^4(\tau)} \int_{\tau_i}^{\tau} d\tau' a^5(\tau') D(\tau') \right)\, , \nonumber
\end{eqnarray}
where we have assumed that the backreaction $D(\tau')$ vanishes for sufficiently early times $\tau' < \tau_i$.  We emphasize that 
\eqref{eqn4} holds irrespective of whether a fluid approximation applies. 

Further consequences of \eqref{eqn4} are explored in Appendix~\ref{appa}. In particular, we show that if backreaction arises
in a purely radiative sector, the $D$-dependent contribution to 
$\bar\varepsilon(\tau)$  in eq.\ \eqref{eqn4} cancels exactly the second term on the right hand side, so that $H(\tau)$ is in this sence $D$-independent. 
For other equations of state, however, the $D$-dependence does not cancel in eq.\ \eqref{eqn4}, \emph{i.e.}, the Friedmann equation is
modified. 

\section{Linearized Inhomogeneitites and backreaction in cosmological fluids} 
\label{sec3}
In this section, we first consider cosmological matter fields in dynamical scenarios for which fluid dynamics applies and where
inhomogeneities can be followed by linearized fluid equations. 
The backreaction formalism of section~\ref{sec2} can then be elaborated on in detail. 

\subsection{The backreaction term for a fluid}
For the case that matter fields can be described as non-ideal fluids with finite pressure
and/or non-vanishing shear- and bulk viscosities $\bar\zeta$ and $\bar\eta$, the backreaction term $D(a)$ 
can be written explicitly as~\cite{Floerchinger:2014jsa} 
\begin{eqnarray}
D &=&-\frac{1}{a} \int d^{3} q P_{\theta p}(\vec{q}) \nonumber \\
&& + \frac{1}{a^{2}}\left(\bar{\zeta}+\frac{4}{3} \bar{\eta}\right) \int d^{3} q P_{\theta \theta}(\vec{q})  
\nonumber \\
&& +\frac{1}{a^{2}} \bar{\eta} \int d^{3} q\left(P_{w}\right)_{j j}(\vec{q})\, .
\label{eqn5}
 \end{eqnarray}
Here, $\left\langle\tilde{\theta}\left(\vec{q}_{1}\right) \tilde{\theta}\left(\vec{q}_{2}\right) \right\rangle =\delta^{(3)}\left(\vec{q}_{1}+\vec{q}_{2}\right) P_{\theta \theta}\left(\vec{q}_{1}\right)$ defines the power spectrum $P_{\theta \theta}$ in terms of a spatial average
$\langle  \dots \rangle$ of the expansion scalar 
$\theta(x) = \int d^{3}q\, {\tilde\theta}(q) e^{i q x}$. The spectra for vorticity $w$ and pressure $p$ are defined analogously. 

To understand how to arrive at \eqref{eqn5}, one may consider the case of a fluid with pressure but with vanishing viscosities. The fluid dynamic equation for the energy density takes then the form 
\begin{equation}
\dot \varepsilon + \vec v \cdot \vec \nabla \varepsilon +  (\varepsilon + p) \left(3 \frac{\dot a}{a} + \vec \nabla \cdot \vec v \right) =0\, . 
\label{eqn6}
\end{equation}
Here, the flow field $u^\mu = (\gamma, \gamma \vec{v})$ with $\gamma = 1/a\sqrt{1-v^2}$ allows for small inhomogeneities 
$| \vec{v}| \ll 1$ on top of the Hubble flow. The combination 
$\vec v \cdot \vec \nabla \varepsilon +  \varepsilon \,  \vec \nabla \cdot \vec v $ is a total derivative with 
vanishing spatial average and therefore, 
\begin{eqnarray}
\tfrac{1}{a}\dot{\bar \varepsilon} + 3 H \, (\bar \varepsilon + \bar p) =
\tfrac{1}{a}\langle  \vec v \cdot \vec \nabla p \rangle
=-\tfrac{1}{a}\langle \delta p\,  \vec \nabla  \cdot \vec v  \rangle\, .
\label{eqn7}
\end{eqnarray}
The spatial average $\langle \delta p\,  \vec \nabla  \cdot \vec v  \rangle$ over second order perturbations can then be 
expressed in terms of an integral over the power spectrum $P_{\theta p}(\vec{q})$. This is the first line of \eqref{eqn5}. 
One obtains the last two lines in \eqref{eqn5} by including the shear and bulk viscous contributions in the Navier-Stokes approximation in eq.\ \eqref{eqn6} and
following the same derivation. 

\subsection{Non-ideal fluids with standard growing modes}
\label{sec3ba}
The main aim of the present section is to calculate $D$ explicitly for model scenarios of large-scale structure formation 
in which cosmological matter is treated in the fluid approximation. To this end, we expose first some qualitative considerations which 
inform our choice of model scenarios:

The growth of large scale structure after photon decoupling is thought to follow a growing mode approximately $\propto a$. 
This growing mode is realized, \emph{e.g.}, when cold dark matter (CDM) is described as an ideal fluid. However, 
the term $D$ in \eqref{eqn5} vanishes for a pressure-less and non-viscous ideal fluid. 
On the other hand, one-component fluids with pressure and/or viscosity are known to show a modified growth of 
cosmological large scale structure \cite{Li:2009mf, Velten:2011bg, Gagnon:2011id, viscous, Barbosa:2017ojt}. For instance, for a one-component fluid with pressure,  the Jeans criterion implies 
that over-densities oscillate rather than grow if their mass is below the Jeans mass. This shows that it is not
easy to generate any backreaction without modifying the late-time power-law growth of cosmological large scale structure.

In building cosmological models with backreaction, one bold approach could be to start without taking into account any constraint from the approximately linear growth of large scale structure; one may hope that at least some of the models constructed this way evade existing constraints or indicate modifications that will be supported by future observations. Alternatively, a very conservative approach would be to restrict the study to cosmological models which show a linear growing mode at all scales and which are therefore designed to pass observational 
constraints on large scale structure. In the following, we adopt this conservative approach. 

Two-component fluids in which a fluid with pressure and/or viscosity is coupled gravitationally to a pressure-less fluid show a standard linear growing mode in linear perturbation theory. One well-known case is a fluid with pressure coupled to a pressure-less fluid. This
is realized, \emph{e.g.}, in the $\Lambda$CDM model where baryonic matter retains after recombination 
a small ionization and thus a small pressure. It is then the gravitational coupling to pressure-less cold dark matter that 
ensures the growth $\propto a$ of the density contrast of baryonic matter on all scales, while the baryonic pressure does some
work during the same epoch. We calculate the  very small  backreaction effect for this 
$\Lambda$CDM scenario in section~\ref{sec3a} before investigating other two-component models that allow for somewhat
 larger backreaction effects.

\subsection{Gravitationally coupled two-component fluids}
\label{sec3ca}
For a one-component fluid with pressure and finite viscosity, the equations of motion for the expansion scalar $\theta_\bk$
and the energy density contrast $\delta_\bk \equiv  \delta \varepsilon_\bk / \exb$ read at linear order and for subhorizon fluctuations 
\begin{eqnarray}
\dot{\delta}_{\bk}&=&-(1+w)\theta_{\bk}-3\Hc (c^2_s-w) \delta_{\bk}\, ,
\label{eq9} \\
\dot{\theta}_{\bk}&=&-(1-3c^2_{ad})\Hc \theta_{\bk}
+k^2\Psi_{\bk}
\nonumber \\
&&+\frac{c^2_s}{1+w}k^2\delta_{\bk}
-\frac{4}{3} k^2 \frac{\eta}{(1+w)\exb a}  \theta_{\bk}\, ,
\label{eq10} \\
k^2 \Phi_{\bk}&=&-\frac{3}{2}\Hc^2 \delta_{\bk} \, ,
\label{eq11}  \\
k^2 \Psi_{\bk}&=& -\frac{3}{2}\Hc^2 \left(
\delta_{\bk} 
+4 \frac{\eta}{(1+w)\exb a} \theta_{\bk} \right) \, .
\label{eq12}
\end{eqnarray}
Here, the subscript $\bk$ denotes Fourier modes of wavelength $2\pi/k$   and eqs.~\eqref{eq11} and \eqref{eq12} are the Poisson
equations for the two Newtonian potentials $\Phi_{\bk}$ and $\Psi_{\bk}$.  
We have parametrized  the equation of state  in terms of 
\begin{eqnarray}
w&=&\frac{\bar p}{\bar \varepsilon}\, ,
~~
c^2_s = \frac{dp}{d\varepsilon}\, ,
~~
c^2_{ad}=\frac{\dot{\bar p}}{\dot{\bar \varepsilon}}
=w-\frac{\dot{w}}{3(1+w)}\Hc\, ,\qquad
\label{eq13}  
\end{eqnarray}
where $H=\Hc/a$ and where $c_s$ denotes the velocity of sound.
We assume in the following that $w$, $c_s$, $c_{ad}$, $\eta\Hc/(\exb a)$ are much
smaller than 1, so that we can set them equal to 0, apart from terms in which they are multiplied by
$k^2$. In this approximation, the shear viscous difference between the two Newtonian potentials
is neglected, $\Phi_{\bk}=\Psi_{\bk}$.  

We now consider two sectors: one with pressure and/or shear viscosity that gives rise to a 
backreaction $D$ (we denote it therefore by a subscript $D$),  and one which is standard pressure-less and non-viscous 
cold dark matter (subscript $C$). The spatially averaged energy densities of both sectors evolve independent of each other, 
but it is the sum of their gravitational attractions that enters the trace of Einstein's field equations and that 
determines the expansion history. 
This is seen in the background equations 
\begin{eqnarray}
&&\frac{1}{a}\dot{\bar{\varepsilon}}_D  + 3 H \, \bar{\varepsilon}_D =D\, ,
\label{eq14}\\
&&\frac{1}{a}\dot{\bar{\varepsilon}}_C + 3 H \, \bar{\varepsilon}_C =0\, ,
\label{eq15}\\
&& \frac{\ddot a}{a^3} = \frac{1}{a} \dot H + 2 H^2 = \frac{4\pi G_\text{N}}{3} 
(\bar{\varepsilon}_D+\bar{\varepsilon}_C) \, .
\label{eq16}
\end{eqnarray}
We denote matter inhomogeneities in the two sectors by ${\delta}_{\bk}$, ${\theta}_{\bk}$ and
${d}_{\bk}$, ${\vartheta}_{\bk}$, respectively. To linear order, these inhomogeneities are coupled
gravitationally via the Poisson equation,\footnote{For non-viscous and pressure-less cold dark matter, the dynamics at 
the mildly non-linear scales of baryon-accoustic oscillations can be described by matching to a non-ideal fluid description 
where the effective viscosity and pressure are parametrically small $O(\Hc^2/k_m^2)$, with 
$k_m \sim 1 \frac{\rm h}{\rm Mpc}$ being the matching scale~\cite{viscous,Floerchinger:2016hja}. Here, we neglect these small contributions although they can be as large as the smallest effects invoked in the following discussion.} 
\begin{eqnarray}
\dot{\delta}_{\bk}&=&-\theta_{\bk}\, ,
\label{eq17} \\
\dot{\theta}_{\bk}&=&-\Hc \theta_{\bk}+k^2\Psi_{\bk}
+ c^2_s k^2\delta_{\bk}
-\frac{4}{3} k^2 \frac{\eta}{\exb_D a}  \theta_{\bk}\, ,
\label{eq18}\\
\dot{d}_{\bk}&=&-\vartheta_{\bk}\, ,
\label{eq19} \\
\dot{\vartheta}_{\bk}&=&-\Hc \vartheta_{\bk}+k^2\Psi_{\bk}\, ,
\label{eq20}\\
k^2 \Psi_{\bk}&=&-\frac{3}{2}\Hc^2 \left(
\Omega_C d_\bk +\Omega_D\delta_{\bk} \right)   \, ,
\label{eq21} 
\end{eqnarray}
where $\Omega_D \equiv \varepsilon_D/(\varepsilon_D + \varepsilon_C)$ is the energy fraction of the 
sector with pressure and/or viscosity, and $\Omega_D+\Omega_C=1$. Within this section, we consider
a matter-dominated Einstein Universe with $\Hc\sim a^{-1/2}$  and approximately constant energy fractions 
$\Omega_D$, $\Omega_C$. To linear order in inhomogeneities,
this two-component system does not develop vorticity and with
$\delta p_\bk = c_s^2\, {\delta \varepsilon_D}_\bk = c_s^2\, \bar\varepsilon_D\, \delta_\bk$, its
backreaction reads 
\begin{equation}
D  =  
- \frac{1}{a} \exb_D \int d^3 q \, c^2_s P_{\theta \delta}(q) 
+ \frac{1}{a} \exb_D  \int d^3 q \, 
\frac{4}{3}\frac{\eta}{\exb_D a}  P_{\theta\theta} (q) \, .
\label{eq22}
\end{equation}
It is helpful to rewrite the evolution of the linear perturbations \eqref{eq17}-\eqref{eq21} in the standard 
two-component fields
\begin{equation}
	\left( \begin{array}{cc} \phi_1\\ \phi_2	\end{array} \right)
		\equiv  \left( \begin{array}{cc} \delta_\bk \\  \frac{-\theta_\bk}{\Hc}	\end{array} \right)\, ,\qquad
	\left( \begin{array}{cc} \phi_3\\ \phi_4	\end{array} \right)
		\equiv  \left( \begin{array}{cc} d_\bk \\  \frac{-\vartheta_\bk}{\Hc}	\end{array} \right)\, .
		\label{eq23}
\end{equation}
One finds
\begin{eqnarray}
\phi'_1&=&\phib\, ,
\label{eq24} \\
\phi'_2&=&- \frac{1}{2} \phib+\frac{3}{2}
\left(\Omega_D \phia +(1-\Omega_D) \phic \right)
\nonumber \\
&& -c^2_s \, a \, 
 \tilde{k}^2 \phia
-c^2_\eta \, a\,
 \tilde{k}^2 \phib\, ,
\label{eq25}\\
\phi'_3&=&\phid\, ,
\label{eq26} \\
\phi'_4&=&-\frac{1}{2}\phid+\frac{3}{2}
\left(\Omega_D \phia +(1-\Omega_D) \phic \right)\, ,
\label{eq27}
\end{eqnarray}
where the prime denotes a derivative with respect to $\ln a$, and where we used
$( 1+\Hc'/\Hc) = 1/2$ for an Einstein Universe. 
In eq.~\eqref{eq25}, the velocity of
sound enters in the combination
$c_s^2\tfrac{k^2}{\Hc^2} \phi_1 = c_s^2\tfrac{\Hc_0^2}{\Hc^2} \tfrac{k^2}{\Hc_0^2} \phi_1 =  c_s^2\, a\, \tilde{k}^2  \phi_1$,
where we used $\Hc_0^2/\Hc^2  = a$ for an Einstein Universe and 
\begin{equation} 
 	\tilde{k} \equiv \frac{k}{\Hc_0}\, .
	\label{eq28}
\end{equation}
Similarly, the viscous corrections are written in terms of
\begin{equation}
	c^2_\eta= \frac{4 \eta \Hc}{3\exb_D a}\, .
	\label{eq29}
\end{equation}
In the absence of pressure and viscosity, $c_s^2=c_\eta^2 = 0$, the evolution equations have
the growing mode  $\phia = \phib = \phic = \phid \propto a$.  In more general two fluid models discussed below, this 
generalizes to
\begin{equation}
	\phia=\phib\propto a \, ,\qquad      \phic=\phid\propto a\, .
	\label{eq30}
\end{equation}
As a consequence of the relative minus sign between the growth of $\delta_\bk$ and the growth of $\theta_\bk$, 
it is then clear that the pressure-induced backreaction term in the first line of \eqref{eqn5} is always positive if evaluated on the
growing mode. This can be made more explicit by rewriting 
\begin{equation}
D  =  
\Omega_D \exb H \left(  \int d^3 q \,c^2_s \, P_{\phia \phib}(q) 
+  \int d^3 q \, 
c^2_\eta \,  P_{\phib \phib} (q) \right) \, .
\label{eq31}
\end{equation}

\subsection{Baryons coupled to CDM}
\label{sec3a}
The solutions of the linear evolution equations \eqref{eq24}-\eqref{eq27} depend on the material properties
$c_s^2$ and $c_\eta^2$ and their scale-dependence. In this subsection, we consider first the
case of normal baryonic matter gravitationally coupled to cold dark matter (CDM) shortly after the era of recombination. 
The square of the baryon sound velocity is~\cite{Ma:1995ey}
\begin{equation}
c_{s}^{2}=\frac{\dot{P}_{b}}{\dot{\rho}_{b}}=\frac{k_{\mathrm{B}} T_{b}}{\mu_{\rm mol}}\left(1-\frac{1}{3} \frac{d \ln T_{b}}{d \ln a}\right)\, ,
\label{eq32}
\end{equation}
where $\mu_{\rm mol}$ is the mean molecular weight (including electrons and all ions of H and He), $T_b$ is the temperature of the 
baryon fluid and $k_{\mathrm{B}}$ is Boltzmann's constant. For a rough estimate, we use 
$\mu_{\rm mol} = 1\, \mathrm{g}/{\mathrm{mol}}$ 
and we assume a matter-dominated universe with $T_b = T_\circ\, a^{-1}$ and $T_\circ = 2.7\, {\mathrm{K}}$. This yields
$c_{s}^{2}\, a = \tfrac{4\, k_{\mathrm{B}} T_\circ}{3\mu_{\rm mol}} = 3\cdot 10^{-13}$ which corresponds to $c_s \approx 2\cdot 10^{-5}$ at recombination ($a \approx 0.001$). 
Consistent with the approximations  leading to \eqref{eq9} -- \eqref{eq12}, this sound velocity is
small enough to be neglected in the equations of motion except where $c_s^2$ is enhanced by a factor $\tilde{k}^2$. 
The shear viscous contribution $\propto c_\eta^2$ of the baryonic fluid is much smaller, so that we can set $c_\eta^2 = 0$ 
in the following. 

Since $T_{\mathrm{b}} \sim a^{-1}$ in a matter-dominated Universe, eq.~\eqref{eq32} allows us to 
set in the evolution equations \eqref{eq24}-\eqref{eq27}
\begin{equation}
c_s^2 a \kt^2= \frac{3\alpha}{2}\, ,
\label{eq33}
\end{equation}
where $\alpha$ is constant in time but depends on the wave-number $k$.
The resulting system has been considered
in chapter 8.3 of Weinberg's Cosmology book~\cite{Weinberg:2008zzc}. For the ansatz 
$\phi_i = \phi_{\circ,i} \left(a/a_\circ\right)^\mu$, $i = 1,2,3,4$, $\phi_1 = \zeta\, \phi_3$, 
equations~\eqref{eq24}-\eqref{eq27} lead to 
\begin{eqnarray}
\xi \left( \mu^2+\frac{\mu}{2}+\frac{3\alpha}{2} \right)&=&\mu^2+\frac{\mu}{2}\, ,
\label{eq34} \\
\mu^2+\frac{\mu}{2}&=&\frac{3}{2}+\Omega_B\xi-\frac{3}{2}\Omega_B\, .
\label{eq35} 
\end{eqnarray}
Here, we have renamed the energy fraction $\Omega_D$ in \eqref{eq24}-\eqref{eq27} by $\Omega_B$, 
since it is the baryonic fluid that exhibits pressure in the example of this subsection. For $\Omega_B \ll 1$, the 
growing-mode solution of \eqref{eq35} is $\mu=1$, \emph{i.e.}, the small pressure contribution leaves the
$a$-dependent growth of the density contrast (almost) unaffected, while the amplitude of the pressure-full 
fluid component is somewhat reduced by~\cite{Weinberg:2008zzc}
\begin{equation}
	\phi_1 = \xi\, \phi_3\, ,\quad \hbox{\rm with}\quad \xi = \frac{1}{1+\alpha}\, .
	\label{eq36}
\end{equation}
To calculate from this solution the backreaction 
\eqref{eq31}, we need to know the power spectrum $P_{\phia \phib}(q)$. Since $\phib = \phia^\prime = \phia$ on the
growing mode, we have $P_{\phia \phib}(q) = P_{\phia \phia}(q) = P_{\delta \delta}(q) $. 
The solution~\eqref{eq36} relates
the energy density fluctuations of the baryonic fluid to
those of cold dark matter,  $\delta_{\bf k} = \xi\, d_{\bf k}$, and therefore $P_{\delta \delta}(q) = \xi^2\, P_{dd}(q) $.
The power spectrum  
$P_{dd}({\bf k})$ of cold dark matter
is given by  $\left\langle d_{\bf k_1}\,  d_{\bf k_2}  \right\rangle =\delta^{(3)}\left( {\bf k_1}+{\bf k_2}  \right) 
P_{dd}\left({\bf k_1}\right)$. 
A very simple parametrization that is sufficient for our purposes is
%
\begin{eqnarray}
P_{dd}(q) &=& {\cal N}\, a^2\, 
\left[\left(\frac{q}{q_{\mathrm{eq}}}\right) \Theta\left(q_{\mathrm{eq}}-q\right)  \right.
\nonumber \\
&& \qquad \left.+\left(\frac{q}{q_{\mathrm{eq}}}\right)^{-3} \Theta\left(q-q_{\mathrm{eq}}\right)\right]\, ,
\label{eq37}
\end{eqnarray}
with $q_{\mathrm{eq}} = 0.02\,  \tfrac{\mathrm{h}}{\mathrm{Mpc}}$ and ${\cal N} = 10^2
\left(\tfrac{\mathrm{Mpc}}{\mathrm{h}}\right)^3$. This allows us to write \eqref{eq31} as
\begin{equation}
D  =  
\Omega_B \exb H\,c^2_s  \int d^3 q  \, \frac{P_{dd}(q) }{\left( 1 + \tfrac{2}{3} c_s^2\, a\, \tfrac{q^2}{\Hc_0^2}\right)^2} \, ,
\label{eq38}
\end{equation}
where the denominator of the integrand arises from the factor $\xi^2$ in $P_{\delta \delta}(q) =  \xi^2(q)\, P_{dd}(q) $. 
This factor $\xi^2$ is scale-independent but it depends on the wavenumber $q$, see eq.~\eqref{eq33}. This is important: without 
the factor $\xi^2(q)$, the integral \eqref{eq38} would be logarithmically 
UV-divergent. The pressure-induced correction  $\xi^2(q)$ is the physics effect that regulates this
UV-divergence.  To evaluate \eqref{eq38}, we introduce the shorthand
\begin{equation}
	A = \frac{2}{3} c_s^2\, a\, \frac{q_{\mathrm{eq}}^2}{\Hc_0^2} 
	\, , \label{eq39}
\end{equation}
and we find
\begin{eqnarray}
	D  &=&  
\Omega_B \exb H\,c^2_s\, a^2\, {\cal N} q_{\mathrm{eq}}^3\, 4\pi {\cal F}(A) 
\nonumber\\
&=&  
6\pi \, \Omega_B \exb H\, {\cal N} q_{\mathrm{eq}}\, \Hc_0^2\, a\, A\, {\cal F}(A) 
\label{eq40}
\end{eqnarray}
with
\begin{eqnarray}
	{\cal F}(A) &=&  \int x^2\, dx\, \frac{x\Theta(1-x) + x^{-3}\Theta(x-1)}{\left(1+A\, x^2\right)^2}
		\nonumber \\
		&=& \frac{A\left( -1+A\ln\left[1+\tfrac{1}{A}\right]\right) + \ln\left[1+A\right]}{2 A^2}\, .
		\label{eq41}
\end{eqnarray}
For the Hubble constant $\Hc_0 = \tfrac{1}{3000} \tfrac{\mathrm{h}}{\mathrm{Mpc}}$, the prefactor in eq.~\eqref{eq40} is 
$6\pi\, {\cal N} q_{\mathrm{eq}}\, \Hc_0^2 = 4.2\cdot 10^{-6}$ and $A=7.9 \cdot 10^{-10}$. This  
yields ${\cal F}(A) = 10.2$ and 
\begin{equation}
	D =\lambda \, \exb_B \, H\,  a\qquad \hbox{with}\quad  \lambda \approx 3.4\cdot 10^{-14}\, .
	\label{eq42}
\end{equation}
Given the smallness of the pre-factor $\lambda$, this backreaction is negligible and cannot affect the evolution 
\eqref{eq14} of cosmological background fields in a measurable way. 

\subsection{An explicit example of a modified Friedmann equation}
\label{sec3b}
How would the background fields and the Friedmann equation be modified in the presence of a more sizeable backreaction 
of the form \eqref{eq42}? As this would not be the physically realized baryonic matter, we switch notation $\exb_B \to \exb_D$ in this section. Inserting $D =\lambda \, \exb_D \, H\,  a$ into \eqref{eq14}, we find  with $\tfrac{d}{d\tau} = H\, a^2 \tfrac{d}{da}$ the equation of motion
\begin{equation}
	a \frac{d}{d a} \bar{\varepsilon}_D+3 \bar{\varepsilon}_D=\lambda a \bar{\varepsilon}_D\, ,
	\label{eq43}
\end{equation}
which has the analytic solution
\begin{equation}
	\bar{\varepsilon}_D=\bar{\varepsilon}_{\mathrm{i},D}\left(\frac{a_i}{a}\right)^{3} 
	e^{\lambda\, \left(a-a_i\right)}\, .
	\label{eq44}
\end{equation}
Here, $\bar{\varepsilon}_{\mathrm{i},D}$ is the energy density at the initial time $\tau_i$ at which 
$a_i=a(\tau_i)$. Since $a \leq 1$, the factor $e^{\lambda\, \left(a-a_i\right)}$ is $O(1)$ throughout the evolution
and the correction to the characteristic power-law decay of cold dark matter becomes negligible for very small $\lambda$.
The factor $e^{\lambda\, \left(a-a_i\right)}$ is always larger than unity which is consistent with the general expectation 
that the work done by backreaction increases the energy density of the corresponding matter component. 

In the modified Friedmann equation \eqref{eqn4}, the energy density \eqref{eq44} enters in the combination
\begin{eqnarray}
	&&\bar{\varepsilon}_D - \frac{1}{a^{4}(\tau)} \int_{\tau_{i}}^{\tau} d \tau^{\prime} a^{5}\left(\tau^{\prime}\right) D\left(\tau^{\prime}\right)
	\nonumber \\
	&&= \bar{\varepsilon}_D - 	\frac{\lambda}{a^{4}} \int_{a_{i}}^a d a^{\prime} {a^\prime}^4   \bar{\varepsilon}_D(a^\prime) 
	\nonumber \\
	&&= \bar{\varepsilon}_{\mathrm{i},D}\left(\frac{a_i}{a}\right)^{3} 
		\left(  \frac{e^{\lambda\, \left(a-a_i\right)}\ - 1 + \lambda\, a_i}{\lambda\, a} \right)\, .
		 \label{eq45}
\end{eqnarray} 
For $a\gg a_i$, this allows us to write \eqref{eqn4} in the form
\begin{eqnarray}
	H^2(\tau) 
	&=& \frac{8\pi G_N}{3}
	\left(\Omega_C+\Omega_D \frac{e^{\lambda\, a }\ - 1 }{\lambda\, a}   \right) 
	\left( \bar\varepsilon(\tau)\vert_{\lambda=0}\right)\, .\qquad  \label{eq46}
\end{eqnarray}
Compared to the standard Friedmann equation $H^2 =  \tfrac{8\pi G_N}{3}\, 
\left(\Omega_C+\Omega_D\right) \bar\varepsilon\vert_{\lambda=0}$, backreaction thus increases 
Hubble's constant ($\tfrac{d}{d\lambda} H^2(\tau) > 0$).
Starting from $a \tfrac{d}{d a} \bar{\varepsilon}_D+3 (1+w) \bar{\varepsilon}_D=\lambda a \bar{\varepsilon}_D$ instead of 
eq.~\eqref{eq43}, one can obtain analytic results also for a warm dark matter component ($\bar p=w\bar \varepsilon$) 
supplemented by a backreaction
of the form \eqref{eq42}. For all $w < \tfrac{1}{3}$, backreaction leads to an increase of $H^2(\tau)$, see
appendix~\ref{appa}.
%
\begin{figure}[t]
\includegraphics[width=0.45\textwidth]{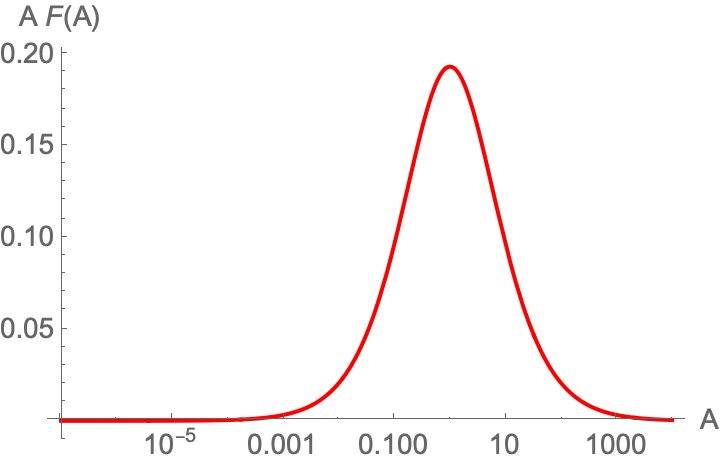}
\caption{The function $A\, {\cal F}(A)$ defined in \eqref{eq41} that sets the scale of the backreaction. 
}
\label{fig1}
\end{figure}
%
\subsection{Two-component fluids with increased backreaction}
\label{sec3c}
For baryons gravitationally coupled to cold dark matter, we had found in section ~\ref{sec3a} a truly minute backreaction of
$O\left(10^{-13} \Omega_B \bar\varepsilon H\right)$. To what extent could backreaction in gravitationally coupled 
two-component fluids be larger if the material properties were different from those of standard
baryonic matter? 

\subsubsection{Backreaction as a function of $c_s^2$ for $c_s^2\, a = {\rm const.}$}
\label{sec3c1}
To address this question, let us consider first a two-component system of the kind described in section ~\ref{sec3a}, but where
the sound velocity $c_s^2$ of the pressure-full matter component takes a value different from that of the baryonic fluid. Since the backreaction $D$ in \eqref{eq31} is proportional to $c_s^2$, one may try to increase the backreaction $D$ by increasing $c_s^2$.
However, there is a tight upper bound to any such effort: the amplitude $\xi$ of the growing mode in \eqref{eq36} is reduced by increasing
$c_s^2$. This reduction is larger for higher wavenumber $q$ and it thus regulates the integral \eqref{eq38} that determines $D$. 
The resulting $c_s^2$-dependence  of $D$ is encoded in the
dependence of the factor $A\, {\cal F}(A)$  in \eqref{eq40}. Plotting this factor in Fig.~\ref{fig1}, we see that backreaction reaches 
its maximum for $A \approx 1$ which corresponds to $c_s^2\, a = 4\cdot 10^{-4}$ and that it decreases for larger $c_s^2$. 
The value $c_s^2\, a = 4\cdot 10^{-4}$ still lies
within the range of validity of our approximation scheme ($c_s^2 \ll 1$) in which only terms proportional to $c_s^2 k^2$ were kept. 

For the baryonic fluid, we
had determined $A\, {\cal F}(A) = 8.1\cdot 10^{-9}$. According to Fig.~\ref{fig1}, ${\rm max}\left[A\, {\cal F}(A)\right] = 0.2$ 
is a factor $2.5\cdot 10^7$ larger. 
Therefore, a dark matter component with minimal pressure ($c_s^2 \ll 1$) that couples gravitationally to 
cold dark matter and that participates in the growth of large scale structure with standard growing mode could 
exhibit a backreaction term as large as
\begin{equation}
D \approx 10^{-6}\, \Omega_D \bar\varepsilon H\,  a\, .
\label{eq47}
\end{equation} 
While this is much larger than the value  \eqref{eq42} obtained for the baryonic fluid, 
it still falls short of being detectable.

\subsubsection{Backreaction for pressure and viscosity with arbitrary scale dependence}
\label{sec3c2}
So far, we have considered only dark matter components which exhibit negligible viscosity and whose sound velocity has
a particular $a$-dependence $c_s^2\, a = {\rm const}$. We next ask to what extent other material properties could yield
backreactions that are even larger than eq.~\eqref{eq47}. To this end, we explore now cosmological two-component fluids
in which one of the two components shows material properties of arbitrary scale $\kx_s $, $\kx_\eta$ and
with arbitrary power-law scale-dependence
\begin{eqnarray}
c^2_s a &=&\kx_s a^\nus\, ,
\label{eq48} \\
c^2_\eta a=\frac{4}{3} \frac{\eta \Hc}{\exb } &=&\kx_\eta a^\nuv\, .
\label{eq49}
\end{eqnarray}
We introduce the shorthands $\phi_i \equiv \frac{\phi_{in}}{a_{in}}\phit_{i}$  for $i=1,2,3,4$.  Eqs. (\ref{eq24})-(\ref{eq27}) can be
 cast into two second-order differential equations for the growth of the density contrasts $\phit_{1}$ and $\phit_{3}$, 
\begin{eqnarray}
a^2 \frac{d^2\phit_1}{da^2}
+\left(\frac{3}{2}+\kx_\eta \kt^2 a^{\nu_\eta} \right)a \frac{d\phit_1}{da} &&
\label{eq50} \\
-\left(\frac{3}{2}\Omega_D-\kx_s \kt^2 a^{\nu_s} \right)\phit_1
&=&\frac{3}{2}(1-\Omega_D)\phit_3\, ,
\nonumber\\
a^2 \frac{d^2\phit_3}{da^2}
+\frac{3}{2}a \frac{d\phit_3}{da}
-\frac{3}{2}(1-\Omega_D)\phit_3
&=&\frac{3}{2}\Omega_D\phit_1\, .
\label{eq51} 
 \end{eqnarray}
We consider again the case $\Omega_D\ll 1$ in which the growing mode is (almost) unperturbed.  
In the limit $\Omega_D\to 0$, eq. (\ref{eq51}) has the solution $\phit_3=a$,
so that eq. (\ref{eq50}) becomes
\begin{equation}
a^2 \frac{d^2\phit_1}{da^2}
+\left(\frac{3}{2}+\kx_\eta \kt^2 a^{\nu_\eta} \right)a \frac{d\phit_1}{da}
+\kx_s \kt^2 a^{\nu_s}\phit_1
=\frac{3}{2}a.
\label{eq52} 
\end{equation}
For the viscosity-free case ($\kx_\eta=0$) with constant $c_s^2 a$ (i.e. $\nu_s=0$), the solution $\phit_1=\xi\, a$
of this equation reduces to eq.~ \eqref{eq36}. A numerical solution of the general case~\eqref{eq52}
is possible. We find it more instructive, however, to consider two special cases that illustrate the generic properties
of the solution.

\underline{Case I:  $\kx_s=0$, $\kx_\eta =$ finite.}
For this case, it is convenient to use the evolution equation $\phit_2=\phit'_1=a\, d\phit_1/da$ and to 
rewrite eq.~\eqref{eq52},
\begin{equation}
a\frac{d \phit_2}{d a}+\frac{1}{2}\phit_2 
+\kx_\eta \kt^2 a^{\nu_\eta} \phit_2=\frac{3}{2}a.
\label{eq53} 
\end{equation}
The solution is
\begin{eqnarray}
&&\phit_{2}(a)=e^{\frac{\kx_\eta \kt^2}{\nu_\eta}(a_i^{\nu_\eta}-a^{\nu_\eta})} \sqrt{\frac{a_i}{a}}\phit_{2i}
\nonumber \\
&&\quad +\frac{3}{2\nu_\eta}a^{-\frac{1}{2}}
\left(-\frac{\nu_\eta}{\kx_\eta \kt^2} \right)^{\frac{3}{2\nu_\eta}} 
e^{-\frac{\kx_\eta \kt^2}{\nu_\eta}a^{\nu_\eta}}
\nonumber \\
&&\quad \times
\left(\Gamma\left[\frac{3}{2\nu_\eta},-\frac{\kx_\eta  \kt^2 a_i^{\nu_\eta}}{\nu_\eta} \right] 
-\Gamma\left[\frac{3}{2\nu_\eta},-\frac{4\kx_\eta  \kt^2 a^{\nu_\eta}}{\nu_\eta} \right]\right).
\nonumber \\
&~&
\label{eq54} 
\end{eqnarray}
The first term in the above expression can be neglected for 
$\phit_{2i}=a_i \ll 1$. 
For $\nu_\eta>0$, a useful approximate solution can be 
obtained if one observes that the
evolution is characterized by two regimes. During the early evolution, 
for small $a$, the last term in the rhs of eq. (\ref{eq53}) is negligible and
the solution is $\phit_2=a$. At late times, for $a\sim 1$, this last term becomes
dominant and the solution is  $\phit_2\simeq 3a^{1-\nu_\eta}/(2\kx_\eta \kt^2)$. 
A good fit of both regimes, giving also the correct order of magnitude for the
short intermediate region, is given by the relation
\begin{equation}
\phit_{2}(a)=\frac{a}{1+\frac{2}{3}\kx_\eta \kt^2 a^{\nu_\eta}}.
\label{eq55}
\end{equation}

\underline{Case II: $\kx_\eta=0$, $\kx_s =$  finite.}
For this case the analytical solution is complicated and not very useful 
because of the presence
of strong oscillations for large $\kt$. 
However, if one averages over the oscillations, 
the average field $\langle \phit_1 \rangle$ can be described by a simple 
expression. In analogy with the viscous case, for small $a$, the solution 
is $\phit_1=a$, while for $a\sim 1$, the last term in the rhs of the
averaged eq. (\ref{eq52})
dominates and  $\langle\phit_1\rangle\simeq 3a^{1-\nu_v}/(2\kx_s \kt^2)$.
For $\nu_s>0$, a relation of sufficient accuracy for our purposes is 
\begin{equation}
\langle \phit_{1} \rangle (a)=\frac{a}{1+\frac{2}{3}\kx_s \kt^2 a^{\nu_s}}.
\label{eq56}
\end{equation}

\begin{figure}[t!]
\centering
$$
\includegraphics[width=0.4\textwidth]{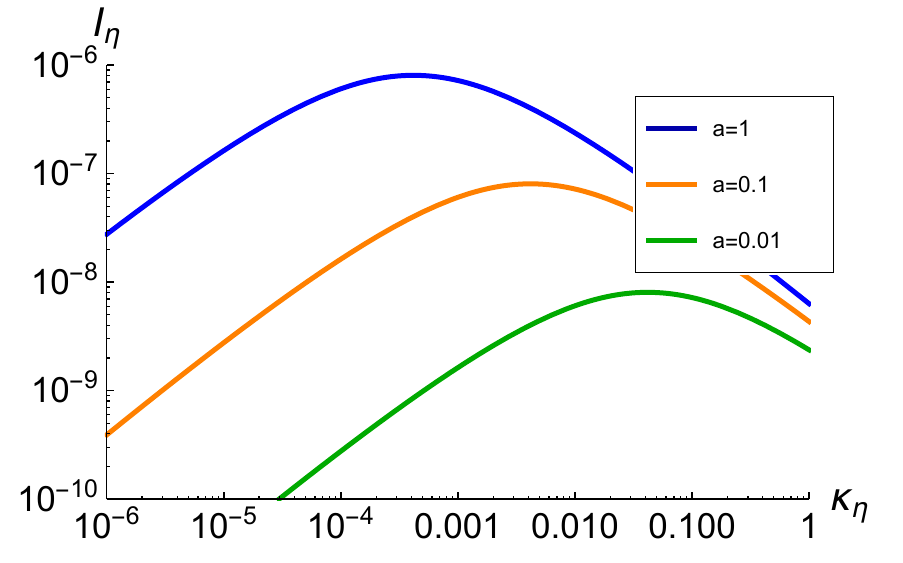}
$$
\caption{The integral $I_\eta$ in \eqref{eq57} determines the relative size of backreaction for a viscous dark matter component
in units of  $\Omega_D \exb H$. The plot is obtained for $c^2_\eta = \kx_\eta a^{\nuv-1}$ with $\nuv = 1$.}
\label{fig2new}
\end{figure}
\subsubsection{Numerical Results}
The denominators of \eqref{eq55} and \eqref{eq56} are scale-dependent generalizations
of the factor $\xi$ in \eqref{eq36} that regulates the UV-divergence of the integral over the power spectrum \eqref{eq38}.
In close analogy to the discussion in subsection~\ref{sec3c1},
this limits the possible growth of backreaction with pressure or viscosity.
To be specific, let us consider for the case I the integral that determines the viscous contribution to the backreaction in \eqref{eq31},
\begin{eqnarray}
	I_\eta (\kx_\eta) &\equiv& \int d^3 q \,c^2_\eta \, P_{\phib \phib}(q) \nonumber\\
	&=& \kappa_\eta a^{\nu_\eta -1} \int d^3 q \frac{P_{dd}(q)}{\left( 1+\frac{2}{3}\kx_\eta \tfrac{q^2}{\Hc_0^2} a^{\nu_\eta}\right)^2}
	\nonumber \\
&=& \left(6\pi {\cal N} \Hc_0^2 \qb_{eq} \right) a\, C\, {\cal F}(C)\, .
\label{eq57}
\end{eqnarray}
Here, the prefactor $6\pi {\cal N} \Hc_0^2 q_{eq}  = 4.2\cdot 10^{-6}$ is the same as in \eqref{eq40}, and the
solution  $\propto C\, {\cal F}(C)$ of the integral is of the same functional form as \eqref{eq40}, but with the $a$-dependent
argument 
\begin{equation}
	C = \frac{2}{3}\kx_\eta \tfrac{q_{eq}^2}{\Hc_0^2} a^{\nu_\eta}\, .
	\label{eq58}
\end{equation}
We know from Fig.~\ref{fig1} that ${\rm max}\left[ C\, {\cal F}(C)\right] \approx 0.2$ and therefore, the viscous backreaction
in \eqref{eq31} has a tight upper bound
\begin{equation}
	D = \Omega_D \exb H \, I_\eta  \leq 10^{-6} \Omega_D \exb H \, a
	\label{eq59}
\end{equation}
for the entire class of  two-component fluid models considered here. This is consistent with the $\kx_\eta$-dependence 
of $I_\eta$ plotted in Fig.~\ref{fig2new}.

\begin{figure}[t!]
\centering
$$
\includegraphics[width=0.4\textwidth]{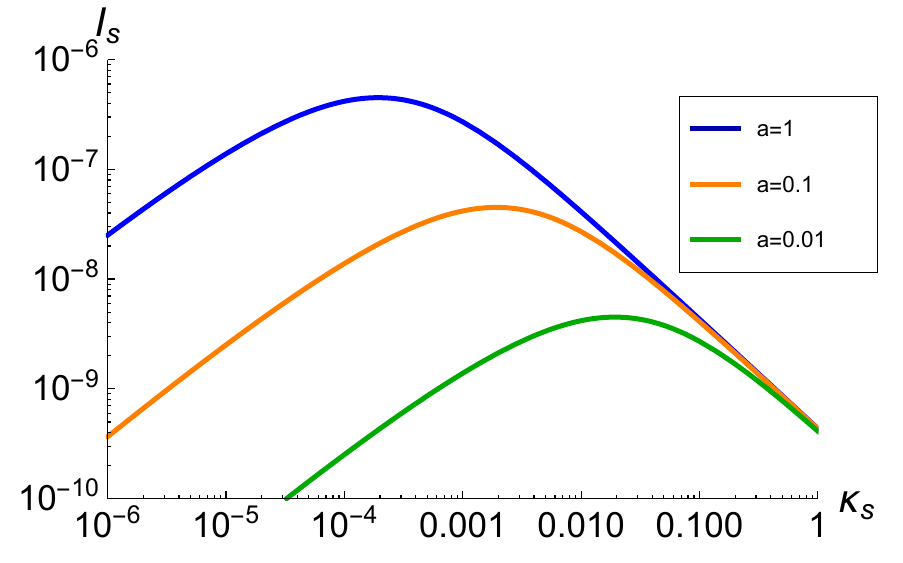}
$$
\caption{The integral $I_s$ in \eqref{eq60} determines in units of  $\Omega_D \exb H$ the relative size of backreaction for a dark 
matter component with sound velocity $c^2_s = \kx_s a^{\nus-1}$. The plot is for $\nus =1$.
}
\label{fig3new}
\end{figure}

We finally write the corresponding contribution to the backreaction~\eqref{eq31} for the case II when viscosity vanishes 
but sound velocity has an arbitrary $a$-dependence, $c^2_s\sim a^{\nu_s-1}$.
\begin{eqnarray}
&&I_{s}(\kx_s) \equiv \int d^3 q \,c^2_s \, P_{\phia \phib}(q)  \nonumber\\
&& \quad =\kx_s a^{\nus-1} \int d^3 q
	\frac{1+\frac{2}{3}(1-\nu_s)\tfrac{ q^2}{\Hc_0^2}  \kx_s\,a^{\nu_s}}{\left( 1+\frac{2}{3}\tfrac{ q^2}{\Hc_0^2} \kx_s a^{\nu_s}\right)^3} {P}_{dd}(q)\, .\qquad
 \label{eq60} 
\end{eqnarray}
While this integral differs somewhat from \eqref{eq57}, the UV-cut on the spectrum ${P}_{dd}(q)$ is set by a similar $q$-dependent
denominator. This explains why the numerical evaluation of \eqref{eq60} in Fig.~\ref{fig3new} yields also in this case a tight upper
bound of the backreaction $D \leq O(10^{-6}) \Omega_D \exb H \, a$.

\section{Backreaction for cosmological matter beyond the linear fluid approximation}
\label{sec4}
Our discussion so far was based on describing matter in terms of ideal or Navier-Stokes fluid dynamics, and on following only scalar perturbations that were assumed to propagate in a linear way.  These assumptions become questionable at late times and on small scales when cosmological structure formation becomes non-linear. Here, we provide a formulation of backreaction that remains valid beyond 
this linearized fluid-dynamic regime and that reduces to the formalism of section~\ref{sec3} in the limit in which Navier-Stokes fluid dynamics applies. 

The need to go beyond a fluid dynamic formulation is particularly clear for the modelling of cold dark matter at late times and sufficiently
small scales. At early times and very large scales, CDM is well-described by the single stream approximation which is equivalent to a description in terms of an ideal and pressure-less fluid. At late times, however, deviations from local equilibrium grow large, and 
the velocity dispersion induced by shell crossing modifies the dynamics \cite{Pueblas:2008uv, Hahn:2014lca, Erschfeld:2018zqg}. In addition to scalar perturbations, also vector perturbations (e. g. vorticity) are generated by non-linear terms. Also the effect of dark matter self-interactions would show up in this regime. At even smaller scales, our current understanding of cosmological evolution is incomplete.

To understand
cosmological evolution, we are particularly interested in the energy density $\varepsilon(x)$. In a particle picture, it receives contributions from rest masses, from interactions and from the kinetic motion of particles (internal energy).  For a one-component fluid with a single (at least approximately) conserved particle quantum number, the differential of this energy density is
\begin{equation}
d\varepsilon = T ds + \mu dn\, ,
\label{eq:differentialEnergy}
\end{equation}
with temperature $T$, entropy density $s$, chemical potential $\mu$ and particle density $n$. Eq.~\eqref{eq:differentialEnergy} 
holds even if the fluid evolution is highly non-linear or turbulent and can be seen as a definition of $T$ and $\mu$. Such more general dynamical scenarios also allow for the definition of 
an entropy current $s^\mu$, as well as a particle number current $N^\mu$, so that one can write
\begin{equation}
s^\mu = s u^\mu + \frac{1}{T}q^\mu- \frac{\mu}{T}\nu^\mu, \quad\quad\quad N^\mu = n u^\mu + \nu^\mu,
\label{eq:defEntropyNumberCurrents}
\end{equation}
with fluid velocity $u^\mu$, heat current $q^\mu$ and diffusion current $\nu^\mu$. The heat current $q^\mu$ and diffusion current $\nu^\mu$ are orthogonal to the fluid velocity, $u_\mu q^\mu = u_\mu \nu^\mu =0$. Depending on the precise definition of the fluid velocity, there is an additional relation for $q^\mu$ and $\nu^\mu$ (e.g.,  $q^\mu = 0$ in the Landau frame or $\nu^\mu = 0$ in the Eckart frame), 
but we keep this frame definition open.

Combining eqs.\ \eqref{eq:differentialEnergy} and \eqref{eq:defEntropyNumberCurrents} and using
$\varepsilon + p = T s + \mu n$, we obtain
\begin{equation}
u^\mu \partial_\mu \varepsilon + (\varepsilon + p) \nabla_\mu u^\mu = \tilde \sigma\, ,
\label{eq:evepsilonTsmn}
\end{equation}
with
\begin{equation}
\tilde \sigma = T \nabla_\rho (s^\rho- q^\rho/T) + \mu \nabla_\rho N^\rho + T \nu^\rho \partial_\rho (\mu/T)\, .
\label{eqn63}
\end{equation}
Entropy production $\nabla_\rho s^\rho$ and the change $\nabla_\rho N^\rho$ of particle number vanish in global equilibrium, 
and so does the heat current $q^\mu$ and the diffusion current $\nu^\mu$. Therefore,
the term $\tilde\sigma$ is proportional to gradients and it vanishes in equilibrium or for an ideal fluid. 

Amongst all possibilities for gradient terms to contribute to \eqref{eq:evepsilonTsmn}, the combination $\nabla_\mu u^\mu$ is distinguished in the cosmological context, because it is non-vanishing already for a homogeneous and isotropic fluid with Hubble expansion. We define therefore
\begin{equation}
\sigma = \tilde \sigma + \pi_\text{bulk} \nabla_\mu u^\mu,
\label{eqn64}
\end{equation}
where $\pi_\text{bulk}$ is defined such that $\sigma$ vanishes for a homogeneous and isotropic but expanding or contracting fluid. Alternatively, and equivalently, it is defined through the tensor decomposition of the energy-momentum tensor with respect to the fluid velocity $u^\mu$,
\begin{equation}
T^{\mu\nu} = \varepsilon u^\mu u^\nu + (p+\pi_\text{bulk}) \Delta^{\mu\nu} + \pi^{\mu\nu} + q^\mu u^\nu + u^\mu q^\nu.
\label{eq:decompositionTmunu}
\end{equation}
Eq.\ \eqref{eq:evepsilonTsmn} becomes then
\begin{equation}
u^\mu \partial_\mu \varepsilon + (\varepsilon + p+ \pi_\text{bulk}) \nabla_\mu u^\mu = \sigma.
\label{eq:evepsilonTsmnbulk}
\end{equation}
By construction, the right hand side of eq.\ \eqref{eq:evepsilonTsmnbulk} vanishes for a homogeneous and isotropic fluid with FLRW expansion. 
In that case, eq.\ \eqref{eq:evepsilonTsmnbulk} gives the standard energy conservation law. 
However, for an inhomogeneous fluid for which homogeneity and isotropy are only symmetries in a statistical sense, 
the right hand side of eq.\ \eqref{eq:evepsilonTsmnbulk} 
can be, and will be, non-vanishing, for example due to terms quadratic in perturbations.

Assuming that metric perturbations are negligible, we can rewrite eq.\ \eqref{eq:evepsilonTsmnbulk}  in terms of deviations
$\vec{v}$ from the local Hubble flow. Using $u^\mu = (\gamma, \gamma \vec v)$, $\gamma = 1/a \sqrt{1-\vec v^2}$, we find
\begin{equation}
\begin{split}
& \dot \varepsilon + \vec v \cdot \vec \nabla \varepsilon +  (\varepsilon + p+ \pi_\text{bulk}) \left(3 \frac{\dot a}{a} + \vec \nabla \cdot \vec v \right) = \frac{\sigma}{\gamma}  . 
\end{split}
\label{eq2}
\end{equation}
As the combination $\vec v \cdot \vec \nabla \varepsilon +  \varepsilon \,  \vec \nabla \cdot \vec v $ is a total derivative with 
vanishing spatial average, the spatial average of eq.\ \eqref{eq2} reads
\begin{equation}
\frac{1}{a}\dot{\bar \varepsilon} + 3 H \, (\bar \varepsilon + \bar p+ \bar \pi_\text{bulk}) = D\, ,
\label{eqb3}
\end{equation}
where
\begin{equation}
D =\frac{1}{a}\langle  \vec v \cdot \vec \nabla (p+\pi_\text{bulk}) \rangle + \left\langle \sigma \sqrt{1-\vec v^2} \right\rangle.
\label{eq:DGeneral}
\end{equation}
We stress again that $D$ vanishes by construction in an exactly homogeneous and isotropic universe and that it starts quadratically in fluctuations. The form of \eqref{eq:DGeneral} is more general than the derivation given in Ref.~\cite{Floerchinger:2014jsa} and used in section~\ref{sec3}, since it does not assume that the fluctuations are small or that $D$ is dominated by terms that are second order
in fluctuations. For illustration, we consider some special cases for this general formula:
\begin{itemize}
\item[(i)] For an ideal fluid with pressure one has $\sigma = \pi_\text{bulk} = 0$, and 
\begin{equation}
D =  -\frac{1}{a}\langle \delta p\,  \vec \nabla  \cdot \vec v  \rangle  .
\label{eq:DIdealFluid}
\end{equation}
This is the case of our illustrative introductory example \eqref{eqn7} that describes a modification of energy density due to work done 
in an imhomogeneous fluid by contraction against local pressure gradients. 
In Fourier-space, $\delta p (x)=\int d^{3}q\, {\delta p}(q) e^{i q x}$, it can be written
as an integral over the power spectrum $D=-\frac{1}{a} \int d^{3} q P_{\theta p}(\vec{q}) $, which is the first term 
of \eqref{eqn5}.
\item[(ii)] In a first order gradient expansion (Navier-Stokes approximation) in the Landau frame  one has $q^\rho=0$ and
the local entropy production
\begin{equation}
\quad\quad \nabla_\mu s^\mu = - \tfrac{1}{T} \left[ \pi_\text{bulk} \nabla_\alpha u^\alpha + \pi^{\alpha\beta} \nabla_{\alpha} u_\beta  \right] - \nu^\alpha \partial_\alpha(\tfrac{\mu}{T}).
\label{eqn71}
\end{equation}
For $\nabla_\mu N^\mu=0$, this leads to the simple expression
\begin{equation}
\sigma = - \pi^{\alpha\beta} \nabla_\alpha u_\beta.
\end{equation}
For exactly homogeneous and isotropic fields without inhomogeneities but with FLRW expansion, the only non-vanishing term 
in $\langle \nabla_\mu s^\mu \rangle$ is proportional to the spatially averaged bulk viscous pressure  
\begin{equation}
\bar\pi_\text{bulk} = - 3 \bar \zeta H\, .
\label{eqn73}
\end{equation}
According to the definition \eqref{eqn64}, this term cancels in $\sigma$ and therefore $D=0$ in the absence of inhomogeneities. 
However, \eqref{eqn73} appears also on the left hand side of \eqref{eqb3}.  The resulting bulk viscous cosmologies without 
inhomogeneities have received much attention in model studies~\cite{Murphy:1973zz,Belinskii,Padmanabhan:1987dg,Fabris:2005ts,Wilson:2006gf,Colistete:2007xi,Mathews:2008hk,Li:2009mf,Velten:2011bg,Gagnon:2011id}. 

Keeping in \eqref{eqn71} terms up to second order in inhomogeneities, one recovers the dissipative terms derived in~\cite{Floerchinger:2014jsa} and given in the second and third line of  eq.~\eqref{eqn5},
\begin{eqnarray}
 D &=& \frac{1}{a^{2}}\left(\bar{\zeta}+\frac{4}{3} \bar{\eta}\right) \int d^{3} q P_{\theta \theta}(\vec{q})  
 \nonumber \\
 && +\frac{1}{a^{2}} \bar{\eta} \int d^{3} q\left(P_{w}\right)_{j j}(\vec{q})\, .
 \label{eq8}
\end{eqnarray}
These terms correspond to the contribution to energy evolution from the entropy production (dissipation) in fluid velocity perturbations. 
\item[(iii)] The backreaction term $D$ in eq.\ \eqref{eq:DGeneral} can also be evaluated in far-from-equilibrium situations. One particularly explicit example is given by a system without conserved particle number, $N^\mu=0$, in the Landau frame $q^\mu = 0$, when \eqref{eqn63}
reduces to 
\begin{equation}
	\tilde \sigma = T \nabla_\rho s^\rho\, .
\end{equation}
In this case, the term $\left\langle \sigma \sqrt{1-\vec v^2} \right\rangle$ in \eqref{eq:DGeneral} is the spatial average over
local entropy production times temperature and an inverse gamma-factor. In any kinetic theory description, even far-from-equilibrium, local entropy production $\nabla_\rho s^\rho \geq 0$  follows from Boltzmann's $H$-theorem and is non-vanishing as a result of collisions away from detailed balance. By working in kinetic theory, or in non-equilibrium quantum field theory, one can therefore evaluate $D$ beyond the simple first-order fluid approximation entering eq.\ \eqref{eqn5}. 

\item[(iv)] From the decomposition \eqref{eq:decompositionTmunu} and the covariant conservation law $\nabla_\mu T^{\mu\nu}=0$, one obtains 
the alternative relation 
\begin{equation}
\sigma = - \pi^{\alpha\beta} \nabla_\alpha u_\beta - \nabla_\alpha q^\alpha + u_\beta u^\alpha \nabla_\alpha q^\beta.
\label{eqn76}
\end{equation}
This is formulated in terms of the energy-momentum tensor only and it does not make reference to entropy. For any form of out-of-equilibrium dynamics that follows the energy momentum tensor and that associates local flow fields (such as e.g. a kinetic theory formulation),  all terms entering \eqref{eqn76} are then given. 

\item[(v)] Relaxing covariant particle number conservation, $\nabla_\mu N^\mu \geq 0$, eq.~\eqref{eq:DGeneral} allows for an increase 
in the energy density by backreaction involving inelastic processes. 
\end{itemize}

These remarks illustrate the general point that working within the framework of a fluid-dynamic evolution with linearized treatment
of inhomogeneities does not exhaust scenarios in which backreaction can occur. We would find it particularly interesting to understand
whether increased backreaction effects can be realized in far-from-equilibrium scenarios that are accompanied by significant entropy
production or significant violations of particle number conservation.

\section{Conclusions}

This work analysed the combined evolution equations for the homogeneous part of the cosmological energy density and the growth of inhomogeneities, accounting for backreaction from the latter to the former.
The evolution equations for the spatial averages (see eqs.\ \eqref{eq1} and \eqref{eq2}) and for the linearized
inhomogeneities (see eqs.\ \eqref{eq9} to \eqref{eq12}) combine to a set of integro-differential equations 
that are difficult to solve in general. 
However, for perturbatively small inhomogeneities, one can solve the linearized evolution equation for perturbations  
in the absence of backreaction, and then
determine the backreaction from the solutions. 
Based on this approximation, we gave in the present paper novel, explicit and compact expressions for how backreaction 
modifies Friedmann's equation in late-time cosmology.  

A basic conclusion that can be drawn is that backreaction on 
the cosmological evolution is not necessarily linked to large metric perturbations, such
as (Post-) Newtonian potentials. The original proposal of Ref.~\cite{Floerchinger:2014jsa} 
had already emphasized this point. We discussed here an explicit realization, 
which has the appealing property that it makes use of two fundamental ingredients of 
the cosmological model: baryonic and dark matter. In section \ref{sec3} we saw that
the growth of structure proceeds in the gravitationally coupled systems of baryons
and dark matter, despite the pressure that develops in the baryonic sector because 
of the small amount of ionization that it retains even after recombination. 
The simultaneous presence of pressure and significant structure growth makes the first 
contribution to the backreaction term of eq. (\ref{eqn5}) nonvanishing. 
The second important conclusion is that the sign of the backreaction term is such
that the effect leads to the enhancement of the averaged energy density that drives 
the expansion. 
For backreaction through pressure this growth arises through the work done
against perturbations. A similar conclusion can be reached for the backreaction through
viscosity. 

In the above example, the known baryon sound velocity is too small for the effect to
be detectable. Replacing baryons by a hypothetical second dark-matter sector with 
nonzero pressure or viscosity enlarges the parameter range and leads to the
enhancement of the effect by several orders of magnitude. However, the standard
cosmological expansion is still not modified significantly. The analysis of 
section \ref{sec3} reveals the reason by providing a refined general picture 
of what limits the 
maximal size of backreaction effects from inhomogeneities of matter fields. 
The backreaction terms  in eq.~\eqref{eqn5} are integrals over 
measured or measurable power spectra. They are
multiplied by the velocity of sound or by a viscous transport 
coefficient respectively. 
One may thus have expected  that the 
maximal size of backreaction is set by the current phenomenological 
constraints on warm dark matter or viscous matter. 
But this is not the case. The growth of large scale structure is certainly 
known to put significant constraints on non-ideal fluid 
alternatives to CDM, but our choice of non-ideal fluids  gravitationally 
coupled to CDM was designed to evade possible constraints 
from LSS. In doing so, we have encountered an even tighter and more generic 
constraint: any pressure or viscous
correction  gives rise to a physical UV-cut-off of the (logarithmically divergent) 
integral over the power spectrum that 
determines backreaction, see eq.~\eqref{eq38}, \eqref{eq57} and \eqref{eq30}. 
This reduces backreaction below the 
conceivably detectable level for arbitrary choices of sound velocity or viscosity.

The analysis in section \ref{sec3} has concentrated on cosmological times and regimes of scales where the evolution of inhomogeneities is close to linear. This allowed to make progress with analytical methods, but also implies -- almost by construction -- that backreaction terms remain small (and perturbation theory remains consistent). On the other side it is also clear that cosmological perturbation theory breaks down at late times and on small scales. The evolution of matter fields is then modified for example by the generation of velocity dispersion generated by stream crossing, or by non-linear terms in the evolution equations for scalar perturbations. Also vector and even tensor perturbations, like vorticity or shear stress, are generated by non-linear terms and can play an interesting role at small scales. Dark-matter self-interactions could modify the dynamics in the non-linear regime as well. Backreaction terms can be written as integrals over equal-time power spectra of inhomogeneities, and it is well possible that these integrals are dominated by small scales where non-linear effects are large. It would therefore be particularly interesting to extend our analysis into that regime. Of course this is technically challenging and needs a good understanding of non-linear physics.

In order to prepare for a future extension into the non-linear and out-of-equilibrium regime, we have formulated the backreaction terms for the evolution of energy density in very general terms in section \ref{sec4}. This new formulation does not rely on the Navier-Stokes fluid approximation and can also be applied for example in the context of far-from-equilibrium quantum field or kinetic theory. It would be interesting to establish in the future an understanding of backreaction effects from the non-linear regime of matter field inhomogeneities to the overall cosmological expansion.

\section*{Acknowledgments} We thank Kfir Blum and Mathias Garny for discussions.
The work of N.~Tetradis was supported by the Hellenic Foundation 
for Research and Innovation (H.F.R.I.) under the  ``First Call for
H.F.R.I. Research Projects to support Faculty members and Researchers and the procurement of high-cost research equipment grant'' (Project Number: 824).
The work of S.~Floerchinger is supported by the Deutsche Forschungsgemeinschaft (DFG, German Research Foundation) under Germany's Excellence Strategy EXC 2181/1 - 390900948 (the Heidelberg STRUCTURES Excellence Cluster), SFB 1225 (ISOQUANT) as well as FL 736/3-1.

\appendix
\section{Further remarks on the modified Friedmann equation}
\label{appa}

In this appendix, we further explore general consequences of the modified Friedmann equation \eqref{eqn4}. To this end, we first define
the critical total energy density $\bar \varepsilon_c$ by
\begin{equation}
H^2 = \frac{8\pi G_\text{N}}{3} \left[ \bar \varepsilon_c - \int_{\tau_\text{I}}^\tau d\tau^\prime \left( \frac{a(\tau^\prime)}{a(\tau)} \right)^4 a(\tau^\prime) D(\tau^\prime) \right] .
\end{equation}
When the integral over $D$ is positive, the total average energy density today $\bar\varepsilon_{0,c }$, for vanishing spatial curvature, is larger than the value usually taken for the critical energy density,
\begin{equation}
\bar\varepsilon_{0,c } > \frac{3 H_0^2 }{8\pi G_\text{N}} .
\end{equation}
If the dissipative term $D$ is non-zero during some time interval, a part of the total background energy density $\bar \varepsilon_c$ has been produced by dissipative processes. It might be useful to separate that part and to decompose the total background energy density additively as
\begin{equation}
\bar \varepsilon_c = \bar \varepsilon + \bar \varepsilon_D
\label{eq:decompositionEnergy}
\end{equation}
where $\bar \varepsilon$ is the conventional contribution of matter and radiation that does not result from backreaction and $ \bar \varepsilon_D$ is the part of the internal  energy that results from the dissipation of perturbations. The part $ \bar \varepsilon_D$ has the evolution
\begin{equation}
\dot{ \bar \varepsilon}_D + 3 \frac{\dot a}{a} (\bar \varepsilon_D + \bar p_D + \bar \pi_{\text{bulk},D}) =  a D ,
\end{equation}
for suitably defined pressure $\bar p_D$ and bulk viscous pressure $\bar \pi_{\text{bulk},D}$. We assume now for simplicity the relation
\begin{equation}
\bar p_D + \bar \pi_{\text{bulk},D} = \hat w_D \, \bar \varepsilon_D,
\end{equation}
where $\hat w_D$ is some constant.  After integrating, one finds
\begin{equation}
\bar \varepsilon_D(\tau) = \int_{\tau_\text{I}}^\tau d\tau^\prime \left( \frac{a(\tau^\prime)}{a(\tau)} \right)^{3+3 \hat w_D} a(\tau^\prime) D(\tau^\prime).
\end{equation} 
Using this, as well as the decomposition \eqref{eq:decompositionEnergy} in the modified Friedmann equation yields
\begin{equation}
\begin{split}
H(\tau)^2 & = \frac{8\pi G_\text{N}}{3} {\bigg [} \bar \varepsilon(\tau) \\
& + \int_{\tau_\text{I}}^\tau d\tau^\prime \left[\left( \tfrac{a(\tau^\prime)}{a(\tau)} \right)^{3+ 3\hat w_D}  - \left( \tfrac{a(\tau^\prime)}{a(\tau)} \right)^4 \right] a(\tau^\prime) D(\tau^\prime) {\bigg ]} .
\end{split}
\label{eq:modifiedFriedmann2}
\end{equation}
One first notes that one recovers the conventional Friedmann equation for $D(\tau^\prime) =0$ or if $\hat w_D = 1/3$. The latter condition would correspond to $\bar \varepsilon_D$ consisting of pure radiation. If $\hat w_D$ is smaller than $1/3$, the Friedmann equation is modified and in particular, today's value of the energy density $\bar \varepsilon$ not produced by backreaction becomes
\begin{equation}
\begin{split}
(\bar \varepsilon)_{0} & = \frac{3 H_0^2}{8\pi G_\text{N}} \\ & -  \int_{\tau_\text{I}}^\tau d\tau^\prime \left[\left( \tfrac{a(\tau^\prime)}{a(\tau)} \right)^{3+ 3\hat w_\text{d}}  - \left( \tfrac{a(\tau^\prime)}{a(\tau)} \right)^4 \right] a(\tau^\prime) D(\tau^\prime),
\end{split}
\end{equation}
which for $D>0$ is smaller than the conventional value because the integral is positive.


\end{document}